\documentclass[useAMS,usenatbib]{./mn2e}
\usepackage{graphicx}
\usepackage{lscape}
\usepackage{times}
\usepackage{amsmath}
\usepackage{amsfonts}
\usepackage{amssymb}
\usepackage{multirow}
\usepackage{longtable}
\usepackage{afterpage}
\usepackage{pdflscape}

\RequirePackage{color}

\title[Comparison of Gaia and asteroseismic distances]{
Comparison of {\it Gaia} and asteroseismic distances
}
\author[M. Y\i ld\i z, Z. \c{C}elik Orhan, S. \"Ortel and M. Roth]{M. Y\i ld\i z$^{1}$\thanks{E-mail:
mutlu.yildiz@ege.edu.tr}, Z. \c{C}elik Orhan$^{1}$, S. \"Ortel$^{1}$ and M. Roth$^{2}$ \\ 
$ $\\
$^{1}$Department of Astronomy and Space Sciences, Science Faculty, Ege University, 35100, Bornova, \.Izmir, Turkey\\
$^{2}$ Kiepenheuer-Institut f\"ur Sonnenphysik, Sch\"oneckstr. 6, D-79104 Freiburg, Germany \\
}

\begin{document}
\date{Accepted 2016 May 15. Received 2016 April 11; in original form 2016 April 11}

\pagerange{\pageref{firstpage}--\pageref{lastpage}} \pubyear{2017}
\def\braket#1{\left<#1\right>}
\newcommand{\yildiz}{Y\i ld\i z }
\newcommand{\etal}{et al. }
\newcommand{\wrt}{with respect to }
\newcommand{\logg}{\log(g) }
\newcommand{\numino}{\mbox{\ifmmode{\overline{\nu_{\rm min}}}\else$\overline{\nu_{\rm min}}$\fi}}
\newcommand{\numin}{\mbox{\ifmmode{\nu_{\rm min}}\else$\nu_{\rm min}$\fi}}
\newcommand{\teff}{\mbox{\ifmmode{T_{\rm eff}}\else$T_{\rm eff}$\fi}}
\newcommand{\teffsun}{\mbox{\ifmmode{{\rm T}_{\rm eff{\sun}}}\else${\rm T}_{\rm eff{\sun}}$\fi}}
\newcommand{\numax}{\mbox{$\nu_{\rm max}$}}
\newcommand{\nuH}{\mbox{\ifmmode{\nu_{\rm minH}}\else$\nu_{\rm minH}$\fi}}
\newcommand{\nuL}{\mbox{\ifmmode{\nu_{\rm minL}}\else$\nu_{\rm minL}$\fi}}
\newcommand{\Dnu}{\mbox{$\Delta \nu$}}
\newcommand{\muHz}{\mbox{$\mu$Hz}}
\newcommand{\kepler}{\mbox{{\it Kepler}}}
\newcommand{\corot}{\mbox{{\it CoRoT}}}
\newcommand{\numaxS}{\mbox{$\nu_{\rm max {\sun}}$}}
\newcommand{\MS}{{\rm M}\ifmmode_{\sun}\else$_{\sun}$~\fi}
\newcommand{\RS}{{\rm R}\ifmmode_{\sun}\else$_{\sun}$~\fi}
\newcommand{\LS}{{\rm L}\ifmmode_{\sun}\else$_{\sun}$~\fi}
\newcommand{\MSbit}{{\rm M}\ifmmode_{\sun}\else$_{\sun}$\fi}
\newcommand{\RSbit}{{\rm R}\ifmmode_{\sun}\else$_{\sun}$\fi}
\newcommand{\LSbit}{{\rm L}\ifmmode_{\sun}\else$_{\sun}$\fi}
\maketitle
\label{firstpage}
\begin{abstract}

Asteroseismology provides  fundamental properties (mass, radius and effective temperature) of solar-like oscillating stars using so-called scaling relations. These properties allow the computation of the asteroseismic distance of stars. We compare the asteroseismic distances with the recently released {\it Gaia} distances for 74 stars studied in Y\i ld\i z et al. There is a very good agreement between these two distances; for 64 of these stars, the difference is less than 10 per cent.  However, a systematic difference is seen if we use the effective temperature obtained by spectroscopic methods; the {\it Gaia} distances are about 5 per cent greater than the asteroseismic distances. 
\end{abstract}


\section{Introduction}

The distance of stars is one of the major astrophysical parameters, whose precise measurement will shed significant light on 
stellar evolutionary theory, and the chemical enrichment of the Milky Way and planetary systems. The {\it Hipparcos}  mission 
({Perryman et al. 1997; van Leeuwen 2007}) allowed a detailed investigation of  stars nearer than 100pc to us. We expect to be able to increase 
this  distance at least 10-fold as a result of the recent {\it Gaia} mission (Gaia Collaboration  et  al.  2016b), with much smaller 
uncertainty.  In this study, the {\it Gaia} distances 
($d_{\rm GAIA}$) are compared with the so-called asteroseismic distances ($d_{\rm sis}$) computed from oscillation frequencies 
of the solar-like oscillating stars studied in \yildiz et al. (2016).

Highly precise parallaxes ($\pi$)  of over one billion stars will be obtained by the ESA {\it Gaia} mission (Gaia  Collaboration  et  al.  2016b). 
The duration of the mission will be at least 5 yr, and the most accurate data on position, proper motion and 
parallax of stars brighter than 20.7 mag will be released in 2022. The mission, launched in 2013, released data (DR1) on  measurements  taken
within  the  first  14  months  of  observations for {more than two million stars} (Gaia Collaboration et al. 2016a) 
which  are  saved in  the Tycho-2 catalogue (H{\o}g et al. 2000).  The astrometric data is available for 74 of 89 target stars in \yildiz et al. (2016).

Asteroseismology is currently experiencing a golden age with the space-based missions {\it Kepler}  (Borucki et al. 2010) and {\it CoRoT} (Baglin et al. 2006 ). {\it TESS} (to be launched in 2018, 
Sullivan et al. 2015) and {\it PLATO} (to be launched in 2025; Catala et al. 2011) will also further the  scientific outcomes in 
this field. Asteroseismology is leading to great advances in the precision of the internal structure model of the solar-like 
oscillating stars in particular (Chaplin \& Miglio 2013). The mass ($M$) and radius ($R$) of these stars can be found from scaling 
relations that relate $M$ and $R$ to the mean of so-called large separation between oscillation frequencies 
($\braket{\Dnu}$), 
frequency of maximum amplitude (\numax) and effective temperature (\teff). The compressibility (the first adiabatic exponent, $\Gamma_{\rm \negthinspace 1s}$) at the surface of 
these stars had until recently been taken as constant in the derivation of these relations (Kjeldsen \& Bedding 1995). However, 
\yildiz et al. (2016) have shown that this is not the case. After obtaining new 
scaling relations based on modifications to conventional ones, they achieved  results in very good agreement with non-asteroseismic predictions, for example, for Procyon A 
(Aufdenberg, Ludwing \& Kervella 2015; Bond et al. 2015).

The radius is required to compute the luminosity of a star.
The modified scaling relation for stellar radius  in solar units (${R_{\rm sca}}/{{\rm R}_{\sun}}$) is given by \yildiz et al. (2016) as
\begin{equation}
\frac{R_{\rm sca}}{{\rm R}_{\sun}}=\frac{(\numax/\nu_{\rm max\sun})}{(\braket{\Delta \nu}/\braket{\Delta \nu_{\sun}})^2}\left( \frac{T_{\rm eff}}{T_{\rm eff{\sun}}}\frac{\Gamma_{\rm \negthinspace 1s{\sun}}}{\Gamma_{\rm \negthinspace 1s}}\right)^{1/2}
\frac{f_{\Delta \nu}^2}{f_{\nu}},
\end{equation}
where $f_{\nu}$ is the ratio of \numax$ $ to acoustic cut-off frequency and $f_{\Delta \nu}$ is defined as {the ratio of $\braket{\Dnu}/\braket{\Dnu_{\sun}}$ to square root of mean density
in solar units ($\sqrt{\rho/\rho_{\sun}}$)}. In conventional scaling relations, $f_{\Delta \nu}$ and $f_{\nu}$ are equal to 1. These are clearly functions of $\Gamma_{\rm \negthinspace 1s}$
(see equations 7 and 11, in \yildiz et al. 2016).

This Letter is organized as follows:
{Section 2 presents the method for the computation of asteroseismic distance.
Section 3 is devoted to the results and their comparison.
Finally, in Section 4, {conclusions are drawn}.

\section{Asteroseismic and {\bf \it G\lowercase{aia}} distances}
In order to determine the distance of a star from distance modulus, its luminosity must be found.
Then, its \teff$ $ and $R$ are required. \teff$ $ can be determined using spectroscopic and photometric methods,
in addition to entirely new asteroseismic methods developed by \yildiz et al. (2014, 2016) for the solar-like
oscillating stars.
However, the most precise prediction of asteroseismology is for radius; $R$ is given in terms 
of \numax$ $, $\braket{\Dnu}$ and \teff. The radii of the target stars are computed from equation (1).
For $\Gamma_{\rm \negthinspace 1s}$, we use the expression given by \yildiz et al. (2016) in terms of \teff:
\begin{equation}
\frac{1}{\Gamma_{\rm \negthinspace 1s}}= 1.6 \left(\frac{\teff}{\teff_{\sun}}-0.96\right)^2+0.607.
\end{equation}

From the luminosity ($L$) of a star, we evaluate its bolometric magnitude.
Bolometric corrections  for the target stars are computed from Lejeune, Cuisinier \& Buser's (1998) tables 
using observed [Fe/H] and \teff$ $ from spectroscopy and asteroseismology. Following this,
we obtain the absolute magnitude from the bolometric magnitude and correction, and 
the distance modulus from the observed visual magnitude ($V$) and absolute magnitude.
In our computations, we take the spectroscopic effective temperature ($T_{\rm eS}$) as \teff$ $ of the target stars.
For testing the effect of \teff$ $ on $d_{\rm sis}$, we also compute $d_{\rm sis0}$ from the asteroseismic effective temperature,
which is computed from the oscillation frequency of min0 (equation 16 in \yildiz et al. 2016).

We directly compute distances from parallaxes given by {\it Gaia}  ($\pi_{\rm GAIA}$) and {\it Hipparcos}  {($\pi_{\rm HIP}$)}.
The {\it Gaia} parallaxes are available for 74 of the stars studied in 
\yildiz et al. (2016).
With the exception of one  red giant {(HD 181907/HIP 95133)}, all are main-sequence and sub-giant stars.

The uncertainty of $d_{\rm sis}$ is computed from the uncertainties in \teff, $R_{\rm sca}$ and $V$. 
Luminosity is perhaps the most uncertain parameter among the fundamental stellar parameters.
{Its uncertainty can be computed from uncertainties of radius and \teff$ $ in quadrature:}
\begin{equation}
\frac{\Delta L}{L}=\sqrt{ \left(2\frac{\Delta R}{R}\right)^2+\left(4 \frac{\Delta \teff}{\teff}\right)^2}.
\end{equation}
The typical uncertainty in asteroseismic distance is computed by taking the luminosity as $L'=L+\Delta L$.
Using $L'$ in place of $L$, we
obtain the most uncertain distance ($d'_{\rm sis}$). The difference between $d'_{\rm sis}$ and $d_{\rm sis}$ ($\Delta d'_{\rm sis}$) is 
the typical uncertainty in asteroseismic distance. We also take into account the uncertainty of $V$ in the computation of $d'_{\rm sis}$. 
{Uncertainty of $\pi$ is computed from $\Delta d'_{\rm sis}$:
\begin{equation}
\frac{\Delta \pi_{\rm sis}}{\pi_{\rm sis}}\approx \frac{\Delta d_{\rm sis}}{d_{\rm sis}},
\end{equation}
}{
where the second and higher terms are neglected. However,
these terms may play a dominating
role when the relative errors are larger than about 10 per cent. }

\section{Results and Discussions}
\begin{figure}
\includegraphics[width=101mm,angle=0]{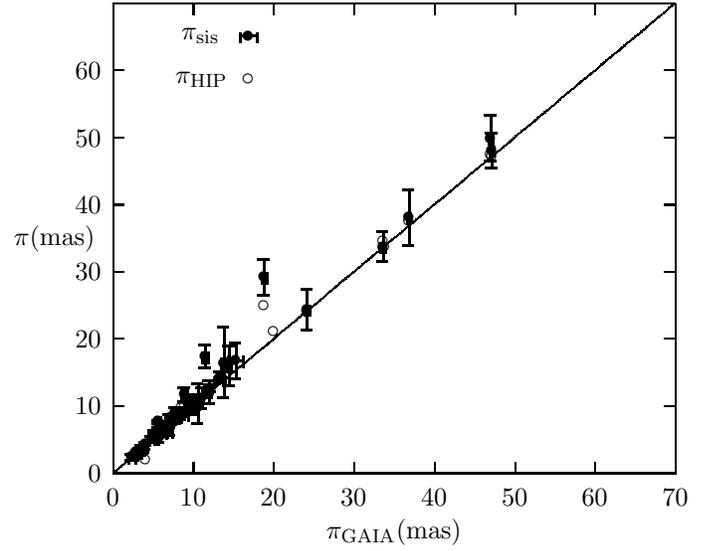}
\caption{  Asteroseismic parallax (filled circle) is plotted \wrt the {\it Gaia} parallax in units of mas. 
For comparison, the {\it Hipparcos} parallax (circle) is also plotted if available.
For one of the stars, namely {KIC 7341231/HIP 92775}, $\pi_{\rm HIP}$ ($d_{\rm HIP}$ = 490 pc) is extremely uncertain but its $\pi_{\rm GAIA}$ and $\pi_{\rm sis}$ are 
in good agreement: $d_{\rm GAIA}=249$ pc,  $d_{\rm sis}=230$ pc.
}
\end{figure}

{ {In Fig. 1, the asteroseismic parallax ($\pi_{\rm sis}$) computed from distance modulus is plotted against to 
$\pi_{\rm GAIA}$}. 
$d_{\rm GAIA}$ of the targets range from 21 pc (16 Cyg A/HIP 96895 and 16 Cyg B/HIP 96901)} to 433 pc {(KIC 10920273/TYC 3547-1968-1)}. 
{For comparison, $\pi_{\rm HIP}$ is} also plotted. {The greatest difference between $\pi_{\rm sis}$ and $\pi_{\rm GAIA}$ occurs for 
KIC 8379927/HIP 97321 ($\pi_{\rm GAIA}=18.76$ mas). Its $\pi_{\rm HIP}$ (24.86 mas) is closer to $\pi_{\rm sis}$ (29.18 mas) than its $\pi_{\rm GAIA}$.} 
There is in general a very good
agreement between $\pi_{\rm sis}$ and $\pi_{\rm GAIA}$ for the majority of the targets. 
However,
there is a systematic difference between $d_{\rm sis}$ and $d_{\rm GAIA}$. The latter is about 5 per cent greater than the former.
{The scatter of the measurement difference is calculated by determining the unit-weight standard deviations, i.e. incorporating 
per measurement the various error contributions. The variance is 2.01 mas$^2$ for $N=74$. Excluding the largest two values the 
variance does not change significantly and is 1.93 mas$^2$.}

The same difference is also seen between $d_{\rm sis0}$ computed from $T_{\rm sis0}$ and $d_{\rm GAIA}$. In Fig. 2, 
the fractional difference between  $d_{\rm sis}$ and $d_{\rm GAIA}$ is plotted with respect to the $V$ magnitude of the target stars.
For most of the targets, the difference between these two distances 
is less than 5 per cent; $\delta d/d_{\rm GAIA}= |d_{\rm sis}-d_{\rm GAIA}|/d_{\rm GAIA}< 0.05$;
however, for five stars, {KIC 1435467/TYC 2666-333-1, KIC 6933899/TYC 3128-1911-1, KIC 8379927/HIP 97321, KIC 9025370/HIP 97321 and KIC 10454113/HIP 97321}, the difference is relatively large. 
With the exclusion of these untypical stars,  the difference slightly increases with respect to $V$ when $V>8.3$ mag, as shown by the solid lines in Fig. 2. 
\begin{figure}
\includegraphics[width=101mm,angle=0]{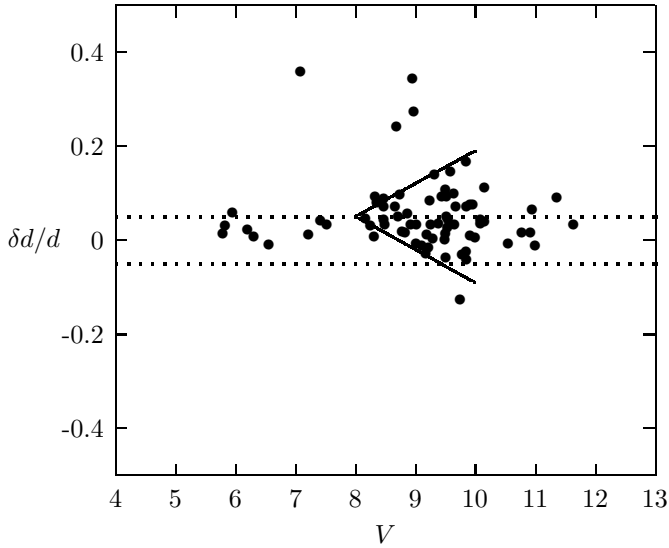}
\caption{Fractional difference between the asteroseismic distance and the {\it Gaia} distances of the targets
is plotted \wrt $V$.
The dotted lines show a 5 per cent uncertainty level. For five stars, the difference is very great. The solid lines 
represent an increase in uncertainty for the stars with $V> 8.3$ mag.
}
\end{figure}

For 64 of 89 stars, the difference between $d_{\rm sis}$ and $d_{\rm GAIA}$ is less than 10 per cent. Silva-Aguirre et al (2012) found 
the difference between $d_{\rm sis}$ and  $d_{\rm HIP}$ for 22 stars less than 20 per cent.

\begin{figure}
\includegraphics[width=101mm,angle=0]{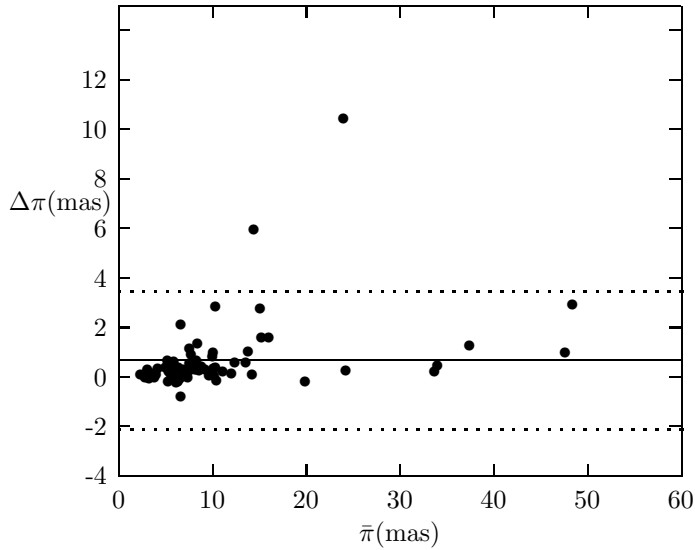}
\caption{
Bland-Altman plot of the difference $\delta \pi= \pi_{\rm sis} - \pi_{\rm GAIA}$ between the asteroseismic and the {\it Gaia} parallaxes of the {\it Kepler}  targets versus the
mean parallax ($\overline{\pi}=(\pi_{\rm sis}+\pi_{\rm GAIA})/2$) of these two measurements.
The solid horizontal {line shows $\overline{\delta \pi}=0.67$ mas}, and the dotted horizontal lines represent  $\overline{\delta \pi}\pm 1.96\rm{std}(\delta \pi)$ with 
{$\rm{std}(\delta \pi)=1.82$ mas for the {\it Hipparcos} parallaxes}.
}
\end{figure}
\begin{figure}
\includegraphics[width=101mm,angle=0]{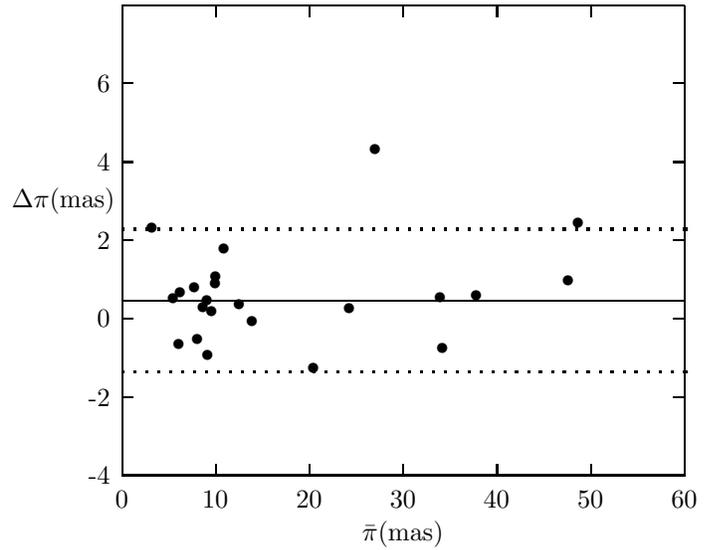}
\caption{ Same as Fig. 3 but the {\it Hipparcos} parallaxes are used in place of the {\it Gaia} parallaxes.
$\overline{\delta \pi}=0.46$ mas and $\rm{std}(\delta \pi) = 3.31 $ mas for the {\it Hipparcos} parallaxes.
}
\end{figure}

If we assume that a small discrepancy is due to an uncertainty in $d_{\rm sis}$, there are two possible reasons: 
uncertainties in $R$ and \teff. 
If due to radius, equation (1) underestimates the radius of stars by about 5 per cent,
and if due to \teff, $T_{\rm eS}$ is found to be 2 per cent less than the \teff$ $ of the targets, or equivalently 120 K less.

If we compare $d_{\rm sis0}$ with $d_{\rm GAIA}$, we find the same mean fractional difference, about 0.05. 
However, the difference between $d_{\rm sis0}$ and $d_{\rm GAIA}$ is less than 10 per cent for 46 of the targets.

{In order to better compare the agreement of the two measurements, Fig. 3 displays the difference $\delta \pi= \pi_{\rm GAIA} - \pi_{\rm sis}$ 
versus the mean out of both measurements $\overline{\pi}=(\pi_{\rm sis}+\pi_{\rm GAIA})/2$ in the form of a Bland--Altman 
diagram (Bland \& Altman 1999).
Based on this scatter plot, we find a positive bias, i.e. the {\it Gaia} measurements give, on average, smaller parallaxes than the seismic measurements.
The mean of the difference is $\delta \pi=0.67$\,mas ($\delta d=6.9$\,pc).
The scatter of the measurement difference $\rm {var}{\delta \pi} = \sum(\delta \pi - \overline{\delta \pi})^2/({N-1})$ is 2.01 mas$^2$ for $N=74$.
Excluding the largest two values, the variance is reduced to 1.93 mas$^2$. Based on the assumption that the distribution is Gaussian, 95\% of the 
values lie between the values $\overline{\pi}\pm 1.96\rm{std}(\delta \pi)=[-2.11,3.45]$ mas for $N=74$ or $[-2.27,3.18]$ mas for $N=72$, respectively.
Given the data the difference and the scatter do not depend on the magnitude of the measurements itself.

Fig. 4 displays a Bland--Altman diagram for comparing $\pi_{\rm sis}$ with the {\it Hipparcos} $\pi_{\rm HIP}$ parallax measurements. Here a bias cannot be detected on the basis of the data set, but there are indications that the scatter, i.e. the variance of the difference, increases with larger distances. However, this finding might still be a result that is due to the small size of the sample.
}

{
The mean difference between  $\pi_{\rm HIP}$ and $\pi_{\rm GAIA}$ is 0.47 mas. This shows that there are very similar differences between three 
different parallaxes. Therefore, it is very difficult to find the sources of uncertainties. However,
uncertainties in \teff$ $ and \numax$ $ might be the main source for the uncertainty of $\pi_{\rm sis}$. 
}

\section{Conclusions}
For solar-like oscillating stars, we can determine $M$, $R$ and even \teff$ $ using 
the new scaling relations given by \yildiz et al. (2016). 
This allows the computation of asteroseismic distances by using bolometric correction and observed $V.$ 
$d_{\rm sis}$ is computed for the stars studied in \yildiz et al. (2016). For 74 of these stars,
the {\it Gaia} distance is also available. There is a very good agreement between $d_{\rm sis}$ and $d_{\rm GAIA}$,
the distances that vary by  
less than 10 per cent for 64 stars.
However, there is a systematic difference between these two distances: $d_{\rm GAIA}$ is 5 per cent greater than $d_{\rm sis}$.
This systematic difference does not depend on the magnitude of the distance.
{Possible sources of uncertainty in $\pi_{\rm sis}$ are due to uncertainties in \teff$ $ and \numax$ $. We hope that more precise {\it Gaia} parallaxes
	to be released in the near future will enlighten the situation. }

	\section*{Acknowledgements}
This work is supported by the Scientific and Technological Research Council of Turkey (T\"UB\.ITAK: 112T989).
MR acknowledges support from the European Research Council under the European Union's Seventh Framework Program 
(FP/2007-2013)/ERC Grant Agreement no. 307117.

\label{lastpage}

\end{document}